\documentclass[11pt]{article}
\usepackage{epsfig}
\usepackage{enumerate}

\def\be{\begin{equation}}
\def\ee{\end{equation}}
\def\bea{\begin{eqnarray}}
\def\eea{\end{eqnarray}}
\newcommand{\lsim}{\mathrel{\mathop{\kern 0pt \rlap
  {\raise.2ex\hbox{$<$}}}
  \lower.9ex\hbox{\kern-.190em $\sim$}}}
\newcommand{\gsim}{\mathrel{\mathop{\kern 0pt \rlap
  {\raise.2ex\hbox{$>$}}}
  \lower.9ex\hbox{\kern-.190em $\sim$}}}

\newcommand{\AmS}{{\protect\the\textfont2
  A\kern-.1667em\lower.5ex\hbox{M}\kern-.125emS}}
\normalsize
\begin{document}

\baselineskip=0.65cm

\begin{center}
\Large
{\bf Comment on ``On an unverified nuclear decay and its role in the DAMA experiment'' (arXiv:1210.5501)}
\vspace{0.5cm}

\rm
\end{center}

\large

\begin{center}

R.\,Bernabei$^{1,2}$,~P.\,Belli$^{2}$,~F.\,Cappella$^{3,4}$,~V.\,Caracciolo$^{5}$,~R.\,Cerulli$^{5}$,
\vspace{1mm}

C.J.\,Dai$^{6}$,~A.\,d'Angelo$^{3,4}$,~A.\,Di Marco$^{1,2}$,~H.L.\,He$^{6}$,~A.\,Incicchitti$^{4}$,
\vspace{1mm}

X.H.\,Ma$^{6}$,~F.\,Montecchia$^{2,7}$,~X.D.\,Sheng$^{6}$,~R.G.\,Wang$^{6}$ and Z.P.\,Ye$^{6,8}$
\vspace{1mm}

\normalsize

\vspace{0.4cm}

$^{1}${\it Dip. di Fisica, Universit\`a di Roma ``Tor Vergata'', I-00133 Rome, Italy}
\vspace{1mm}

$^{2}${\it INFN, sez. Roma ``Tor Vergata'', I-00133 Rome, Italy}
\vspace{1mm}

$^{3}${\it Dip. di Fisica, Universit\`a di Roma ``La Sapienza'', I-00185 Rome, Italy}
\vspace{1mm}

$^{4}${\it INFN, sez. Roma, I-00185 Rome, Italy}
\vspace{1mm}

$^{5}${\it Laboratori Nazionali del Gran Sasso, I.N.F.N., Assergi, Italy}
\vspace{1mm}

$^{6}${\it IHEP, Chinese Academy, P.O. Box 918/3, Beijing 100039, China} 
\vspace{1mm}

$^{7}${\it Laboratorio Sperimentale Policentrico di Ingegneria Medica, Universit\`a
degli Studi di Roma ``Tor Vergata''}
\vspace{1mm}

$^{8}${\it University of Jing Gangshan, Jiangxi, China}
\vspace{1mm}

\end{center}

\normalsize

\begin{abstract}
 We briefly remind references and arguments, already discussed in the past, which confute erroneous claims in 
arXiv:1210.5501. 

\end{abstract}

We are obliged to comment on the paper ``On an unverified nuclear decay and its role in the DAMA experiment''
\cite{scem}, since it contains several erroneous claims, already confuted in the DAMA literature in the past
\cite{perflibra,modlibra,modlibra2,muons12,RNC,ijmd,scineghe09,taupnoz,vulca010,canj11,tipp11}.
We avoid to list here the several arguments which deserve corrections/comments limiting this text to the main point.

The direct decay of $^{40}$K to the ground state of $^{40}$Ar through electron capture 
and the time behaviours of the $^{40}$K counts have
been already quantitatively discussed by DAMA in \cite{scineghe09,taupnoz} and in many conferences. 
Thus, we do not repeat here the many experimental arguments which allow the exclusion of any role 
for $^{40}$K, inviting the reader to read the DAMA literature quoted above.

In particular, a large part of the paper \cite{scem} is dedicated to the electron capture 
to the ground state of $^{40}$Ar (BR$_{EC}$),
whose branching ratio is not well known. Actually, this argument is captious since its contribution to 
the {\it single-hit} events at low energy is only about 10\% of the $^{40}$K total contribution.

The $^{40}$K content of each crystal has been quantitatively determined 
through the investigation of double coincidences \cite{perflibra}.
These values do not depend on BR$_{EC}$.
The measured value of $^{nat}$K content averaged on all the crystals is 13 ppb as reported e.g. 
in \cite{taupnoz}.

Moreover, the authors of \cite{scem} claims that: ``...the presence of the potassium background
poses a challenge to any interpretation of the DAMA results in terms of a Dark Matter model 
with a small modulation fraction. A 10 ppb contamination of natural potassium requires a 20\% 
modulation fraction or more.'' Actually this argument has been already addressed in DAMA literature
as well, as briefly summarized in the following.

By the fact, on the contrary of what is claimed in Ref. \cite{scem}, 
the obtained DAMA model independent evidence
is compatible with a wide set of scenarios regarding the nature of the Dark Matter candidate 
and related astrophysical, nuclear and particle physics.
A few scenarios and parameters (of the many possible) are discussed as examples in
Refs.~\cite{RNC,ijmd,allDM,ijma,epj06,ijma07,chan,wimpele,ldm,bot1,bot2,bot3}, where 
very accurate results on corollary model dependent quests are evaluated in given frameworks 
by applying the maximum likelihood analysis in time and energy of all the events. This procedure 
accounts for all the experimental information carried out by the data, and thus
the information about the counting rate at low energy is correctly considered as a prior.
In particular, as stated in Ref. \cite{taupnoz}, considering the measured $^{40}$K residual 
contamination in the crystals and the remaining background, an upper limit of 0.25 cpd/kg/keV
can be inferred for the unmodulated part of the signal ($S_0$).

Since the measured modulation amplitude ($S_m$) is around $10^{-2}$ cpd/kg/keV, there is no 
reason to claim the necessity to have ``a 20\% modulation fraction or more.''.

Also the procedure to calculate the examples reported in Appendix A of Ref.~\cite{modlibra} 
takes into account the above mentioned constrain on $S_0$.
Therefore, the DAMA model independent evidence
is compatible with a wide set of Dark Matter candidates and scenarios, and hence, for several reasons,
the statement reported in Ref. \cite{scem} 
``A 20 ppb contamination, which is reported as an upper limit by DAMA, disfavors 
any Dark Matter origin of the signal.'' is not correct at all.

\end{document}